# Identification of the growth-limiting step in continuous cultures from initial rates measured in response to substrate-excess conditions

Jason T. Noel, Brenton Cox, Atul Narang[1]
Department of Chemical Engineering, University of Florida, Gainesville, FL 32611-6005

ABSTRACT When steady state chemostat cultures are abruptly exposed to substrate-excess conditions, they exhibit long lags before adjusting to the new environment. The identity of the rate-limiting step for this slow response can be inferred from the initial yields and specific growth rates measured by exposing steady state cultures at various dilution rates to substrate-excess conditions. We measured these parameters for glucose-limited cultures of *E. coli* ML308 growing at various dilution rates between 0.03 and 0.6 1/hr. In all the cases, the initial yields were 20-30% less than the steady state yields. The decline of the yield implies that overflow metabolism is triggered in response to excess glucose. It is therefore unlikely that the initial response of the cells is limited by substrate uptake. The initial specific growth rates of cultures growing at low dilution rates (D = 0.03, 0.05, 0.075, 0.1, 0.3 1/hr) were significantly higher than the steady state specific growth rates. However, the increment in the specific growth rate decreased with the dilution rate, and at D=0.6 1/hr, there was no improvement in the specific growth rate. The initial specific growth rates varied hyperbolically with the dilution, decreasing sharply at dilution rates below 0.1 1/hr and saturating at D=0.6 1/hr. This is consistent with a picture in which the initial response is limited by the activity of glutamate dehydrogenase.

INTRODUCTION

When a steady state chemostat is subjected to an abrupt increase in the flow rate or feed concentration, there is a pronounced overshoot of the substrate concentration that can last several hours or even days (Daigger and Grady 1982, Nielsen and Villadsen 1992). These overshoots often lead to regulatory violations in wastewater treatment plants and product deterioration in industrial bioreactors. In order to develop operating protocols and control strategies for mitigating or preventing these overshoots, it is important to understand their root cause.

The substrate concentration overshoots occur because some physiological (intracellular) process is limiting growth. Indeed, time scale analysis shows that the substrate concentration becomes saturating within a few minutes of increasing the feed flow rate or concentration (Nielsen and Villadsen 1992, Lendenmann and Egli 1995, Gupta et al. 2005). Yet, the specific growth rate does not attain its maximum value for several hours. These submaximal specific growth rates are responsible for the substrate concentration overshoots since they prevent the rapid accumulation of cells required for neutralizing the increased substrate burden. Thus, the key for understanding the overshoots lies in identification of the physiological process limiting initial growth.

---

[1] Email: *narang@che.ufl.edu*

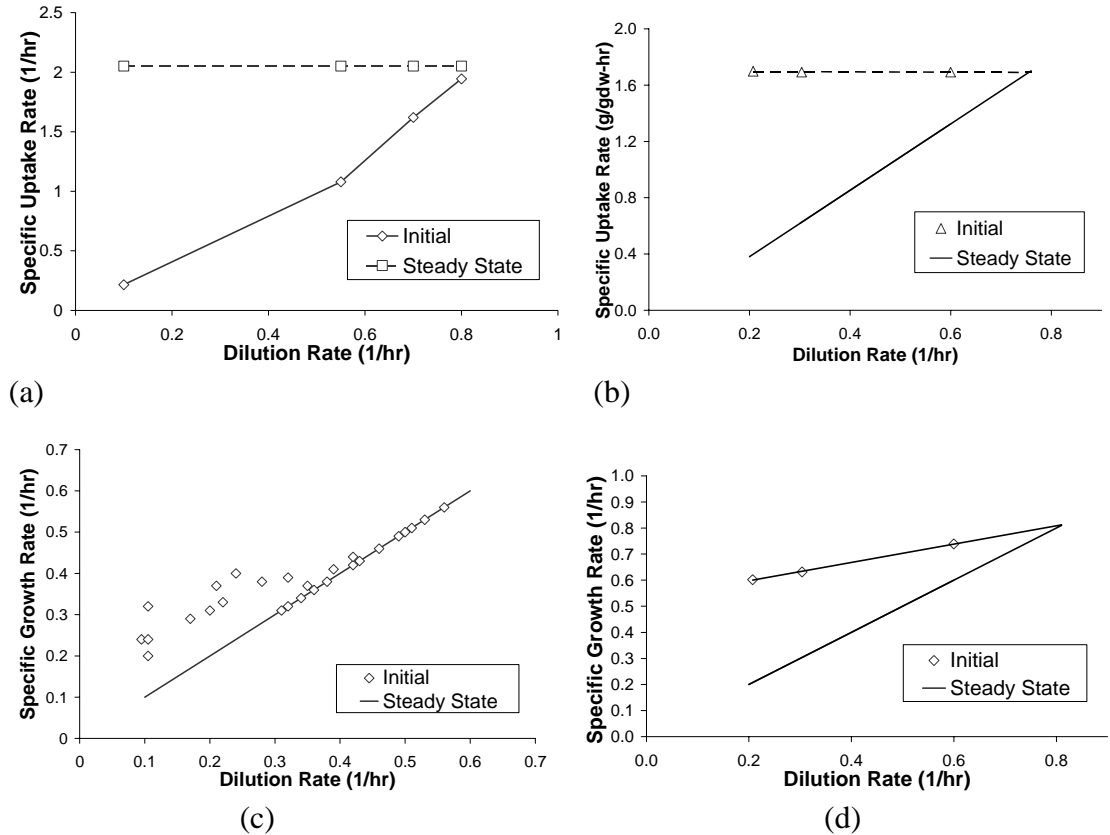

(a) (b)
(c) (d)

**Figure 1:** *Top panel:* Steady state (full lines) and initial (dashed lines) specific substrate uptake rates after a continuous-to-batch shift. (a) Data for glucose-limited *K. aerogenes* (O'Brien et al. 1980). (b) Data for glucose-limited *E. coli* ML308 (Lendenmann and Egli 1995). The maximum specific substrate uptake rate in exponentially growing batch cultures is 2.3 g/gdw-hr. *Bottom panel:* Steady state (full lines) and initial (dashed lines) specific growth rates after a continuous-to-batch shift. (c) Data for glucose-limited *E. coli* B-SG1, a glycogenless mutant of *E.coli* B (Harvey 1970). (d) Data for glucose-limited *E. coli* ML 308 (Lendenmann and Egli 1995). The maximum specific growth rate in exponentially growing batch cultures is 0.92 1/hr.

A convenient experiment for identifying the rate-limiting physiological process is the *continuous-to-batch shift*. In this experiment, steady state cultures drawn from a chemostat are immediately exposed to substrate-excess conditions, and the initial rates of substrate uptake and growth are measured within 10-30 mins of the shift. On this fast time scale, the levels of "slow" variables such as transport/biosynthetic enzymes and ribosomes, which determine the capacity of the cells to import the substrate and transform it into biomass, remain essentially equal to their pre-shift steady state levels (Gupta et al. 2005). Thus, we can gain insight into the identity of the growth-limiting process by comparing the initial rates of substrate uptake and growth with the pre-shift (steady state) levels of the transport/biosynthetic enzymes and ribosomes.

We are aware of three previous studies in which continuous-to-batch shift experiments were performed. These studies showed that when steady state cultures are exposed to substrate-excess conditions
1. The specific substrate uptake rate jumps to maximal or near-maximal levels (Figure 1, top panel). More precisely, O'Brien *et al* observed that regardless of the dilution rate at which the cultures had been growing, the initial specific substrate uptake rate was equal to the *maximum* specific substrate uptake rate, i.e., the steady state specific substrate uptake rate at the washout dilution rate (Figure 1a). On the other hand, Lendenmann & Egli found that the specific substrate uptake rate jumped to near-maximal levels for steady state cultures growing at D<0.6 1/hr (Figure 1b). Although the initial uptake rates were not measured at higher dilution rates, linear extrapolation of the initial and steady state specific uptake rates led them to hypothesize that there is no increase for cultures growing at D>0.75 1/hr.
2. The specific excretion rate jumps dramatically for cultures growing at all the dilution rates (O'Brien et al. 1980). The excreted metabolites consist of primarily acetate, pyruvate, and gluconate (Neijssel and Tempest 1975, Neijssel et al. 1977).
3. The specific growth rate jumps for cultures growing at low dilution rates, but shows no increase for cultures growing at sufficiently high dilution rates (Figure 1, bottom panel). Harvey studied cultures of *E. coli* B-SG1, a glycogenless mutant of *E. coli* B, and found that there was no increase for cultures growing at D>0.3 1/hr (Figure 1c). Lendenmann & Egli observed that the increment in the specific growth rate decreased progressively as the dilution rate was increased from 0.2 to 0.6 1/hr. Based on extrapolation of the initial and steady state specific growth rates, they concluded that there is no increase for cultures growing at D>0.8 1/hr (Figure 1d).

The rapid increase of specific uptake and excretion rates suggests that the initial response is not limited by transport and the central metabolic pathways. Instead, some component of biosynthesis is limiting growth. In the literature, there are two main hypotheses regarding the rate-limiting step in biosynthesis. Some have argued that ribosomes limit growth because under substrate-excess conditions, they become saturated with amino acids, and the specific growth rate cannot be improved until more ribosomes are synthesized (Nielsen and Villadsen 1992). Yet, others have hypothesized that growth is limited by glutamate hydrogenase, the enzyme that catalyzes the synthesis of 85% of the amino acids during carbon-limited growth (Harvey 1970). According to this hypothesis, under substrate-excess conditions, GDH becomes saturated with precursors, so that growth remains sluggish until more GDH is synthesized. As we show below, one can distinguish between these two possibilities by comparing the initial specific growth rates to the steady state levels of GDH and RNA. However, previous studies do not permit such a comparison, since they contain no data for cultures growing below 0.1 1/hr.

In this work, we sought to resolve this deficiency by measuring the initial specific substrate uptake and growth rates for cultures growing over the entire range of dilution rates from 0.03 to 0.6 1/hr. To this end, steady state cultures of *E. coli* ML308 growing at various dilution rates in a glucose-limited chemostat were abruptly exposed to saturating concentrations of glucose. We find that for cultures growing at D>0.1 1/hr, our results are comparable to the data shown in Figures 1b,c,d. However, at lower dilution rates, the initial specific growth and substrate uptake rate decline sharply. Comparison of the initial specific growth rate and the steady state levels of glutamate dehydrogenase and total RNA suggests that the former limits the growth under substrate-excess conditions.

**MATERIALS AND METHODS**

**Strain**

The microorganism used in this study is *Escherichia coli* ML308 (ATCC 15224) obtained from the American Type Culture collection. The strain was preserved at -20C in medium containing 20% glycerol. Stock cultures were prepared from a culture grown on a complex medium (Difco yeast extract).

**Medium**

The medium was prepared with purified water in 20 L polypropylene bottles. It consisted of $KH_2PO_4$ (2.72 g/L), $NH_4Cl$ (0.75 g/L), $EDTA.Na_2.2H_2O$ (82 mg/L), $MgSO_4.7H_2O$ (57 mg/L), $Na_2MoO_4.2H_2O$ (2.4 mg/L), $CaCl_2.2H_2O$ (73.5 mg/L), $FeCl_3.6H_2O$ (2.0 mg/L), $MnCl_2.4H_2O$ (4.9 mg/L), $ZnCl_2$ (1.7 mg/L), $CuCl_2.2H2O$ (9.0 mg/L), $CoCl_2.6H_2O$ (1.2 mg/L). The chemicals purchased were of the highest grade commercially available.

Before sterilizing the medium, it was supplemented with glucose, and the pH was adjusted to 3 with concentrated $H_2SO_4$. The low pH prevents caramelization of the sugar during sterilization and back-contamination of the feed line during operation of the chemostat.

The glucose concentration in the feed was 100 mg/L. At high feed concentrations (~1g/L), the initial specific growth rates measured after continuous-to-batch shifts were significantly lower than the dilution rate at which the cells had been growing in the chemostat. The decline in the initial specific growth rates became evident within a week of inoculating the chemostat, and it was particularly pronounced at high dilution rates. This probably reflects the effect of wall growth (Lendenmann 1994). In the presence of wall growth, the steady state specific growth rate is lower than the dilution rate, since cells shearing off the wall become an additional source of cells inside the bioreactor. We found that at feed concentrations of 100 mg/L, the reactor could be operated for up to 6 weeks without significant wall growth.

**Bioreactor**

The chemostat cultures were grown in a 1.5 L Bioflow III fermenter (New Brunswick Scientific Co.) with a working volume of 1.2 L. The agitation speed was 1000 rpm and the aeration rate was 1.2 L/min. The bioreactor was equipped with automatic pH and temperature control. The pH was maintained at 7.0 by controlled addition of 1M KOH / 1M NaOH and 10% $H_3PO_4$ solutions. The temperature was maintained at 37C. The feed was pumped in by a Masterflex L/S peristaltic pump (7523-70) equipped with a Masterflex EZ Load pump head (7534-04).

The effluent $CO_2$, pH, and temperature were monitored constantly. The effluent $CO_2$ concentration was measured on-line with a $CO_2$ analyzer (Vaisala GMT 222). The $CO_2$ analyzer and the fermenter's pH/temperature monitoring system were connected to a PC for continuous data acquisition.

**Initial specific growth and substrate uptake rates**

To ensure that the chemostat was at steady state, it was maintained at the same condition for at least 20 residence times for all but one of the dilution rates (D=0.03 1/hr in which case only the chemostat was maintained at the same condition for 6 residence times). Before starting the experiment, the maximum specific growth rate (0.91 1/hr) and biomass yield (0.40 gdw/g) of a sample drawn from the chemostat was measured to check if any contaminant or mutant species had invaded the bioreactor.

To measure the initial specific growth and substrate uptake rates after a continuous-to-batch shift, 100 mL of chemostat culture and 60 µL of a 100 g/L glucose stock solution was rapidly added to a 250 mL shake flask that had been prewarmed in an incubated shaker (New Brunswick, Model C24) operated at 37C and 200 rpm. The absorbance at 546 nm was measured every 3 mins by pumping a recycle stream through a spectrophotometer (Spectronic Genesys 10UV). The residence time in the recycle loop was 40 seconds. Samples for sugar measurement were extracted every 5 min for 30 min. To prevent consumption of the sugar during and after the extraction of the sample, the substrate was rapidly separated from the cells by vacuum filtration through a 0.22 µm filter, and stored at -20C.

**Analysis**

The absorbance (546 nm) vs. dry weight calibration curve was generated by measuring the absorbance of dilutions from a stationary phase culture of known cell density. The calibration curve was observed to be linear for optical densities of up to 1.

Glucose concentrations were measured with a Dionex-500 HPLC equipped with an anion exchange column (CarboPac PA10) and a pulsed electrochemical detector. In this method, the sugars are ionized by using a strong base as eluent (10 mM NaOH), separated based on their differential affinity for the anion exchange column, and detected by the pulsed amperometric detector. The manufacturer's protocol resulted in a rapid increase of retention times, presumably due to adsorption of bicarbonate on the column. The addition of 1 mM $Ba(OAc)_2$ to the eluent, as recommended by Cataldi *et al*, precipitated the bicarbonate as $Ba(CO_3)_2$ (Cataldi et al. 1999). The modified eluent dramatically improved the reproducibility and precision. Using this technique, we have measured glucose concentrations down to 10 µg/L.

**RESULTS**

**Determination of the initial specific growth rates and biomass yields**

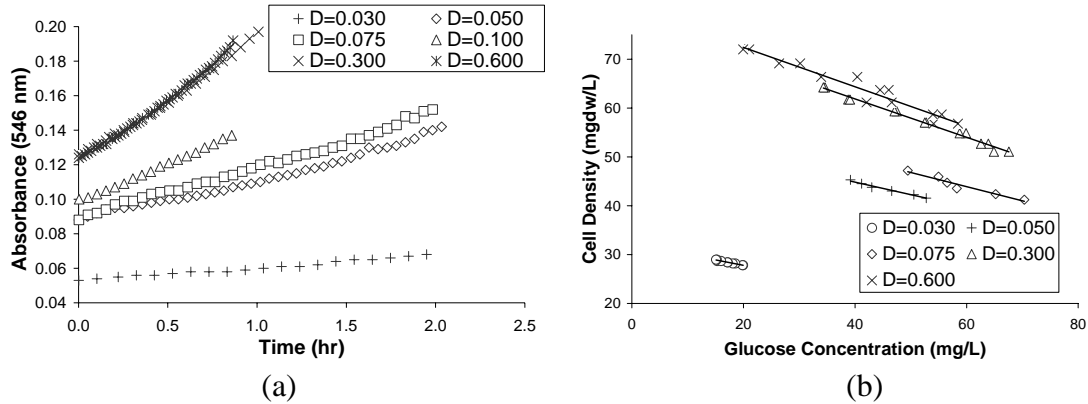

(a)                (b)

Figure 2: (a) The evolution of absorbance after continuous-to-batch shifts at various dilution rates. (b) Cell density vs. glucose concentration after continuous-to-batch shifts at various dilution rates.

The initial specific growth rate and biomass yield at a given dilution rate was determined by allowing the chemostat to reach steady state. A small sample of cells was then drawn from the chemostat, and exposed to excess glucose (20-75 mg/L). The subsequent cell density and glucose concentration were determined every 3 and 5 min, respectively, for 1-2 hr. The experiment was performed at various dilution rates ranging from 0.03 to 0.6 1/hr.

The initial specific growth rates were determined by exponential fits of the absorbance data corresponding to the first 30 min (Figure 2a). For cultures growing at high dilution rates (0.3 and 0.6 1/hr), we observed a marked increase in the specific growth rate after 30 min. For instance, based on the first 30 min of the data, the initial specific growth rate of the culture growing at D=0.3 1/hr was 0.42 1/hr. However, the initial specific growth rate of the same culture based on the data for 30-60 min time period was 0.47. Thus, we calculated the initial specific growth rates based on the data for the first 30 min only.

The initial yields were calculated from linear fits to the cell density vs glucose concentration during the course of the transient (Figure 1b). The linearity of these curves ($R^2 > 0.94$) implies that the initial yield is constant throughout the transient.

The initial specific substrate uptake rate at each dilution rate was calculated from the corresponding initial specific growth rate and yield.

**Variation of initial rates and yield as a function of the dilution rate**

The dashed lines in Figures 3a,b,c show the initial specific uptake rate, specific growth rate, and biomass yield calculated from the data in Figure 2. The full lines in these figures show the corresponding steady state values measured before the culture was exposed to excess glucose. The steady state specific growth rates were assumed to be equal to the prevailing dilution rate. The steady state yields and specific substrate uptake rates were calculated with the formulas, $c/(s_f - s)$ and $D(s_f - s)/c$, respectively, where $s$, $c$ denote the steady state substrate concentration and cell density in the chemostat, $s_f$ denotes the feed concentration of the substrate, and $D$ denotes the dilution rate. At the dilution rates considered in this study, the steady state glucose concentrations are on the order of 0.1 mg/L (Senn et al. 1994), which is negligibly small compared to the feed concentration (100 mg/L). Hence, the steady state yields and specific substrate uptake rates are well approximated by the expressions, $c/s_f$ and $Ds_f/c$, respectively.

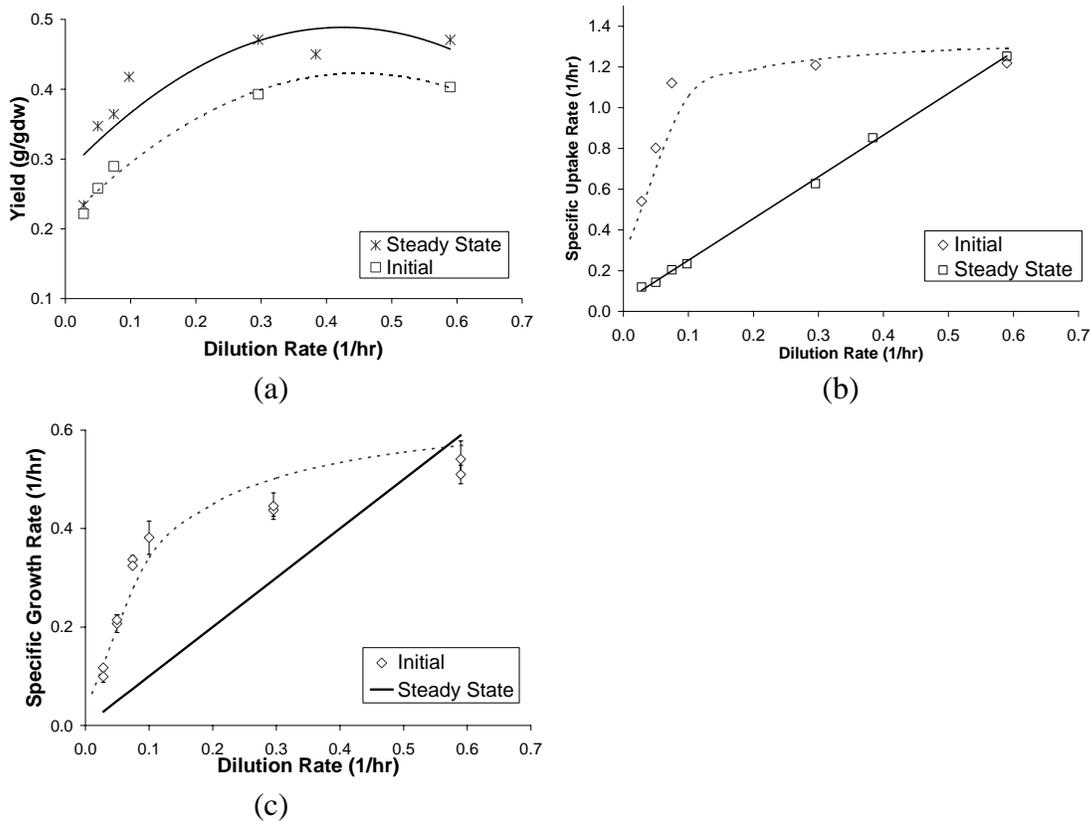

**Figure 3:** Biomass yield, specific substrate uptake rate, and specific growth rate at various dilution rates. The full and dashed lines show the steady state and initial rates, respectively. (a) Yield of biomass. (b) Specific substrate uptake rate (c) Specific growth rate.

Figure 3a shows that at all the dilution rates, the initial yield is 20-30% lower than the corresponding steady state yield. This result is consistent with the results obtained in other studies (Duboc et al. 1998). It implies that high rates of respiration or/and excretion are triggered in response to excess substrate. Thus, it is reasonable to conclude that precursors are in excess, and growth is not limited by the capacity for substrate uptake.

Figure 3b shows that the specific substrate uptake rate increases significantly for cultures growing at all dilution rates below 0.6 1/hr. Thus, the transport enzyme system has excess capacity at these dilution rates. Interestingly, there is no perceptible increase in the specific uptake rate for cultures growing at D=0.6 1/hr. This could be due to two reasons
1. The transport enzyme is already saturated at steady state, so that further increases in exogenous glucose levels produce no improvement in the specific uptake rate.
2. The transport enzyme system is subsaturated at steady state and the specific uptake rate does increase on a fast time scale of 2-3 min, after which it returns to its original steady state value. Since we measured the substrate concentrations every 5 min, this fast transient would not be observed.

The initial yield data implies that the transport cannot be saturated at steady state. If this were true, the addition of glucose to the steady state cultures would produce no physiological change. This is because the cells "see" the environment through the transport enzyme, and if the enzyme is saturated, it will not perceive the addition of glucose to the environment. But Figure 3a shows that the yield clearly decreases in response to excess glucose. This change in the physiological state implies that transport enzymes must be subsaturated at steady state. As we show below, comparison of our data with the data obtained by Lendenmann & Egli indicates that in agreement with the second hypothesis above, the specific uptake rates are indeed higher when measured on a fast time scale.

Figure 3c shows that the initial specific growth rate increases significantly for cultures growing at dilution rates below 0.6 1/hr. No such increase is observed for cultures growing at D=0.6 1/hr. Thus, the initial specific growth rate varies hyperbolically with respect to the dilution rate- it increases sharply in the neighborhood of D=0.1 1/hr and achieves a maximum value of 0.6 1/hr. Importantly, the maximum initial specific growth rate is well below the maximum specific growth rate observed in exponentially growing batch cultures (0.92 1/hr).

## DISCUSSION

The model system in our experiments – growth of *E. coli* ML308 in a glucose-limited chemostat – is identical to that used by Lendenmann & Egli. Comparison with their data, shown in Figures 1b,d, indicates that that the results are in qualitative agreement, but there are quantitative differences. Our initial specific growth and substrate uptake rates are lower than those measured by Lendenmann & Egli. This probably reflects differences in the time scale over which the initial rates were measured. Our initial specific growth and uptake rates are based on data obtained in the first 30 min after exposure to excess glucose. Lendenmann & Egli measured the initial specific growth rates 30 min after the cells had been exposed to excess glucose. We have observed that the specific growth rate increases significantly after the first 30 minutes, particularly at D=0.3 and 0.6 1/hr. Hence, it is not surprising that our initial specific growth rates are lower. Lendenmann & Egli measured the initial specific uptake rates within 7 min of exposure to excess glucose.

It is likely that the specific uptake rates go through a rapid transient in which they pass through a maximum. Such transients could be the consequence of delayed negative feedback. Specifically, if there is a bottleneck at the level of biosynthesis, precursors will accumulate. When the precursor levels become sufficiently high, they inhibit substrate uptake by negative feedback. The time scale of precursor accumulation is ~1 min (Gupta et al. 2005), but it can be delayed substantially by the buffering effect of carbohydrate storage (Sel'kov 1979). Harvey has shown that there is substantial glycogen storage immediately after *E. coli* ML30 is exposed to glucose-excess conditions (Harvey 1970). Thus, it is conceivable that our specific uptake rates are lower probably because they are based on measurements taken over a longer time scale.

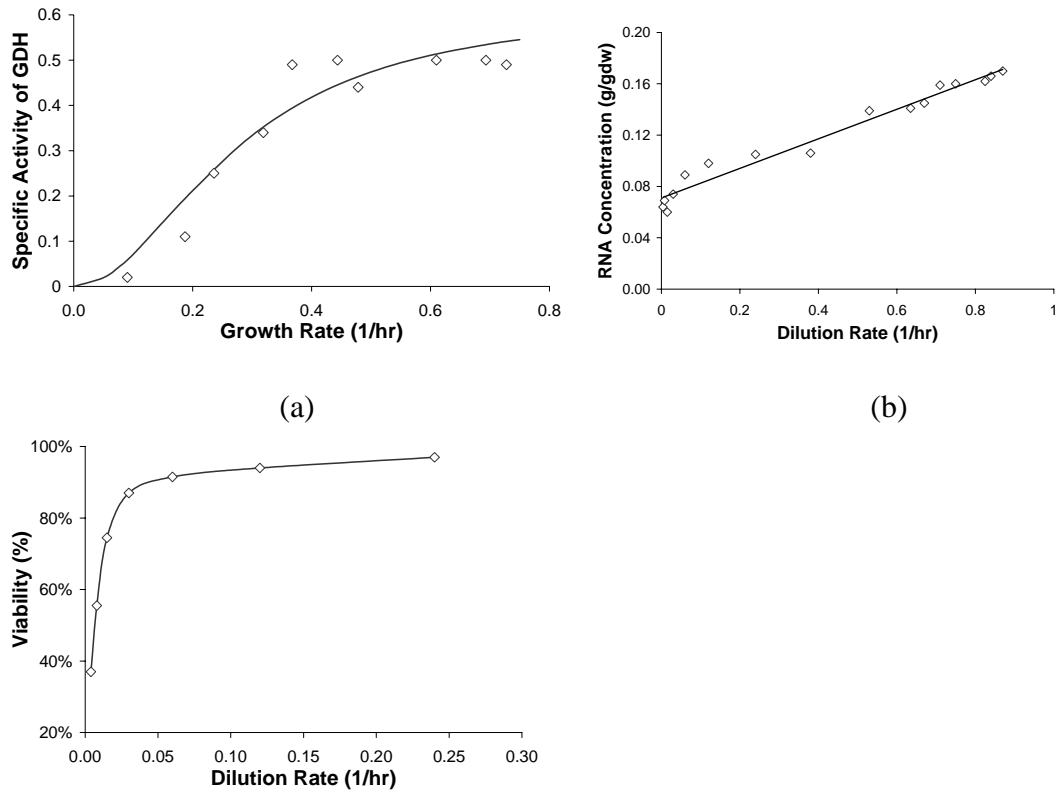

**Figure 4:** Variation of GDH activity and total RNA levels as a function of the dilution rate. (a) GDH activity in *E. coli* W (Senior 1975) (b) Total RNA level in *K. aerogenes* (Tempest et al. 1967) (c) Viability in *K. aerogenes* (Tempest et al. 1967).

As noted above, the decrease of the yield in response to excess glucose suggests that biosynthesis rather than substrate uptake limits the initial response. The key question is which biosynthetic process prevents the cultures from attaining the maximum specific growth rate of 0.92 1/hr. There are two possibilities.
1. Either the activity of GDH, the enzyme that catalyzes the synthesis of ~85% of amino acids during carbon-limited growth, is so low that despite saturating concentrations of precursors, the specific growth rate remains submaximal. In this case, the initial specific growth rate will be proportional to the activity of GDH.

2. Or the activity of GDH and hence, the concentration of amino acids, is sufficiently high, but the level of ribosomes, the organelles that catalyze polymerization of amino acids to proteins, is so low that despite the abundant supply of amino acids, the specific growth remains below maximal levels. In this case, the initial specific growth rate will be proportional to the ribosome level, which, in turn, is proportional to the total RNA level.

Given the fast time scale of the experiment, the GDH and total RNA levels cannot deviate significantly from their pre-shift steady state levels. Thus, we can gain insight into the identity of the growth-limiting process by comparing the variation with respect to dilution rate of the initial specific growth rate and the steady state levels of GDH or total RNA.

Figures 4a,b shows the variation of the steady state levels of GDH and total RNA as a function of the dilution rate. It is evident that the total RNA level increases more or less linearly over the entire range of dilution rates. In fact, it increases two-fold as the dilution increases from 0.2 to the maximum specific growth rate. On the other hand, the initial growth rate increases by no more than 20% over this range of dilution rates. In contrast, the initial specific growth rate and steady state GDH levels are in good agreement. Both increase monotonically at low dilution rates, and saturate at a dilution rate of ca. 0.6 1/hr. To be sure, this correspondence must be viewed with some caution at D<0.1 1/hr since the sharp decline of the initial specific growth rate under such conditions could be due to other variables. For instance, the viability of cells decreases dramatically at low dilution rates (Figure 4c). However, at high dilution rates, it seems plausible that the initial response is limited by GDH levels. This conclusion is corroborated by the fact that the addition of glucose plus amino acids to steady state cultures increases the specific growth rates to values well beyond those observed upon addition of glucose alone (Harvey 1970). This observation is not consistent with a picture in which the initial response is limited by ribosomes due to their saturation with amino acids.

Mathematical models of continuous culture dynamics frequently assume that growth is limited by ribosome levels (Nielsen and Villadsen 1992). However, the initial kinetics of continuous cultures exposed to excess glucose suggest that GDH is limiting. Although this conclusion is supported by our experiments, it must remain tentative until additional experiments are performed to test its implications. This includes measurement of the steady state GDH and total RNA levels for our strain and determination of the specific growth rates in response to amino acid enrichment. These experiments are currently in progress.

**CONCLUSIONS**

We measured the initial yield and specific growth rates of steady state cultures of *E. coli* ML308 in response to excess glucose. We find that
1. At all the dilution rates, the initial yield is lower than the corresponding steady state yield (measured before the culture is exposed to excess glucose). This suggests that biosynthesis rather than uptake limits the initial response of the cultures.

2. The initial specific uptake and growth rates are higher than the corresponding steady state rates at all dilution rates below 0.6 1/hr. At D=0.6 1/hr, neither the specific uptake rate nor the growth rate showed any improvement in response to excess glucose.
3. Comparison of the initial specific growth rate with the profile of the steady state GDH and total RNA levels suggests that biosynthesis of amino acids, catalyzed primarily by GDH, limits the initial response of the cultures.

**Acknowledgements:** We would like to thank Prof. Thomas Egli (EAWAG, Switzerland) for suggestions regarding some of the experimental protocols conducted in this work.

**REFERENCES**

Cataldi, T. R. I., C. Campa, M. A. Angelotti, and S. A. Bufo. 1999. Isocratic separations of closely related mono- and di-saccharides by high-performance anion-exchange chromatography with pulsed amperometric detection using dilute alkaline spiked with barium acetate. J. Chromatography **855**:539--550.
Daigger, G. T., and C. P. L. Grady. 1982. The dynamics of microbial growth on soluble substrates. Wat. Res. **16**:365--382.
Duboc, P., U. von Stockar, and J. Villadsen. 1998. Simple generic model for dynamic experiments with *Saccharomyces cerevisiae* in continuous culture: Decoupling between anabolism and catabolism. Biotechnol. Bioeng. **60**:180--189.
Gupta, S., S. S. Pilyugin, and A. Narang. 2005. The dynamics of single-substrate continuous cultures: The role of ribosomes. J. Theor. Biol. **232**:467--490.
Harvey, R. J. 1970. Metabolic regulation in glucose-limited chemostat cultures of *Escherichia coli*. J. Bacteriol. **104**:698--706.
Lendenmann, U. 1994. Growth Kinetics of *Escherichia coli* with Mixtures of Sugars. Swiss Federal Institute of Technology, Zurich, Switzerland.
Lendenmann, U., and T. Egli. 1995. Is *Escherichia coli* growing in glucose-Limited chemostat culture able to utilize other sugars without lag? Microbiology **141**:71--78.
Neijssel, O. M., S. Hueting, and D. W. Tempest. 1977. Glucose transport capacity is not the rate-limiting step in the growth of some wild-type strains of *Escherichia coli* and *Klebsiella aerogenes* in chemostat culture. FEMS Microbiol. Lett. **2**:1--3.
Neijssel, O. M., and D. W. Tempest. 1975. The regulation of carbohydrate metabolism in *Klebsiella aerogenes* NCTC 418 organisms growing in chemostat culture. Arch. Microbiol. **106**:251--258.
Nielsen, J., and J. Villadsen. 1992. Modeling of microbial kinetics. Chem. Eng. Sc. **47**:4225--4270.
O'Brien, R. W., O. M. Neijssel, and D. W. Tempest. 1980. Glucose phosphoenolpyruvate phosphotransferase activity and glucose uptake rate of *Klebsiella aerogenes* growing in a chemostat culture. J. Gen. Microbiol. **116**:305--314.
Sel'kov, E. E. 1979. The oscillatory basis of cell energy metabolism. Pages 166--174 *in* M. Haken, editor. Pattern Formation by Dynamical Systems and Pattern Recognition. Springer-Verlag, Berlin.

Senior, P. J. 1975. Regulation of nitrogen metabolism in *Escherichia coli* and *Klebsiella aerogenes*: Studies with the continuous culture technique. J. Bacteriol. **123**:407--418.

Senn, H., U. Lendenmann, M. Snozzi, and T. Egli. 1994. The growth of *Escherichia coli* in glucose-limited chemostat cultures: A reexamination of the kinetics. Biochem. Biophys. Acta **1201**:424--436.

Tempest, D. W., D. Herbert, and P. J. Phipps. 1967. Studies on the growth of *Aerobacter aerogenes* at low dilution rates in a chemostat. Pages 240--253 *in* Microbial Physiology and Continuous Culture. HMSO, London.